# Chemical Treatment-Induced Indirect-to-Direct Bandgap Transition in MoS$_2$: Impact on Optical Properties

Yusuf Kerem Bostan[1], Elanur Hut[1], Cem Sanga[2], Nadire Nayir[2,3], Ayse Erol[1], Yue Wang[4] *, and Fahrettin Sarcan[1,4] *

[1]Department of Physics, Faculty of Science, Istanbul University, Vezneciler, Istanbul, 34134, Turkey

[2]Department of Physics Engineering, Istanbul Technical University, Maslak, Istanbul, 34467, Turkey

[3]Paul-Drude-Institute for Solid State Electronics, Leibniz Institute within Forschungsverbund Berlin eV., Hausvogteiplatz 5-7, 10117 Berlin, Germany

[4]School of Physics, Engineering and Technology, University of York, York, YO10 5DD, United Kingdom

**Abstract**

The unique electrical and optical properties of emerging two-dimensional transition metal dichalcogenides (TMDs) present compelling advantages over conventional semiconductors, including Si, Ge, and GaAs. Nevertheless, realising the full potential of TMDs in electronic and optoelectronic devices, such as transistors, light-emitting diodes (LEDs), and photodetectors, is constrained by high contact resistance. This limitation arises from their low intrinsic carrier concentrations and the current insufficiency of doping strategies for atomically thin materials. Notably, chemical treatment with 1,2-dichloroethane (DCE) has been demonstrated as an effective post-growth method to enhance the n-type electrical conductivity of TMDs. Despite the well-documented electrical improvements post-DCE treatment, its effects on optical properties, specifically the retention of optical characteristics and excitonic behaviour, are not yet clearly understood. Here, we systematically investigate the layer- and time-dependent optical effects of DCE on molybdenum disulfide (MoS$_2$) using photoluminescence (PL) spectroscopy and Density Functional Theory (DFT) simulations. Our PL results reveal a rapid reduction in the indirect bandgap transition, with the direct transition remaining unaffected. DFT confirms that chlorine (Cl) atoms bind to sulphur vacancies, creating mid-gap states that facilitate non-radiative recombination, explaining the observed indirect PL suppression. This work demonstrates DCE's utility not only for n-type doping but also for optical band structure engineering in MoS$_2$ by selectively suppressing indirect transitions, potentially opening new avenues for 2D optoelectronic device design.

**Keywords**: Transition metal dichalcogenides, molybdenum disulfide, chlorine doping, density functional theory, Photoluminescence, sulphur vacancies

**Introduction**

Over the past decades, two-dimensional (2D) transition metal dichalcogenides (TMDs) have attracted significant attention due to their ultrathin structures (thickness < 1 nm for monolayers), high carrier mobilities, tuneable bandgaps (in the range of 1-2 eV), and excellent sensing capabilities. [1–4]. Their atomically thin structure leads to high surface-to-volume ratios, making their electrical conductivity highly sensitive to environmental changes - a desirable characteristic for sensing applications [5]. Key examples of TMDs such as WS$_2$, WSe$_2$, MoS$_2$, and MoTe$_2$ are alternative active materials for optoelectronic devices, including LEDs and lasers, and their bandgaps, which can be precisely tuned via thickness, strain and doping [6–10], offering an additional degree of freedom to tailor their properties. A unique feature of layered TMDs is their characteristic transition from an indirect to a direct bandgap in the monolayer limit, attributed to quantum confinement [11]. This is particularly beneficial for optoelectronic devices, as direct bandgaps allow efficient radiative recombination of electrons and holes without requiring phonon assistance, resulting in stronger light emission and absorption. However, despite their remarkable structural, electrical and optical properties, current experimental results show mobilities of ~50 cm²/V·s [12]. This significant discrepancy from theoretical predictions points to dominant scattering mechanisms that limit electrical conductivity, hindering their competitiveness with widely adopted electronic and optoelectronic materials such as Si and GaAs. [13–21]. The successful deployment of TMD materials as active components in high-performance devices critically depends on the availability of both n-type and p-type doping. Due to the strong covalent bonds within each layer and their atomically thin structure, traditional doping techniques used for bulk semiconductors are not suitable for 2D materials, as they often cause significant structural damage or introduce deep-level defects. For instance, ion implantation can create disordered lattices, while high-temperature diffusion doping can lead to interlayer delamination or unintentional phase transitions. These effects severely degrade the optoelectronic performance by reducing carrier mobility, quenching photoluminescence, or increasing trap-assisted recombination [22–25]. On the other hand, due to the atomic thickness of TMDs, their optical and structural properties, as well as carrier dynamics, can be effectively engineered by using various post-growth methods [26–28]. Several prominent post-growth doping techniques have been reported in the literature, notably chemical treatment, ion implantation, plasma doping, thermal annealing, electron beam irradiation, and ultraviolet-ozone treatment. These methods differ in their mechanisms of introducing dopants or modifying carrier concentrations.

Chemical treatment typically relies on surface adsorption or substitutional doping through reaction with precursor molecules. Ion implantation involves accelerating dopant ions into the material, allowing precise control but often causing lattice damage. Plasma doping introduces energetic ions or radicals that can modify the surface or embed dopants. Thermal annealing promotes the diffusion of dopants or activates existing ones by repairing defects. Electron beam irradiation induces defects or modulates local electronic structure through energy transfer, while ultraviolet-ozone treatment introduces oxygen-related species that chemically modify the surface or passivate defects [29–36]. The main challenge in post-growth doping of 2D materials lies in maintaining their excellent optical properties while consistently controlling the doping concentration. In this study, we specifically addressed this challenge in the context of chemical treatment based on 1,2-dichloroethane (DCE).

The chemical formula of 1,2-dichloroethane (DCE) is $C_2H_4Cl_2$. When $MoS_2$ is submerged in DCE solution, chlorine (Cl) atoms interact with $MoS_2$, acting as electron donors and leaving behind ethylene gas ($C_2H_4$). It has been employed as an effective tool for defect engineering and doping in graphene and 2D TMDs. Several studies in the literature have demonstrated the n-type doping of TMDs using 1,2-dichloroethane (DCE) solution. L. Yang et al. reported n-type doping in $WS_2$ and $MoS_2$ via a 12-hour DCE treatment, resulting in reduced contact resistance from ~$10^2$ kΩ·μm to 0.7 kΩ·μm and 0.5 kΩ·μm, respectively [37]. In a recent study by A. Roy et al., it is demonstrated that the Schottky barrier height of $WS_2$ decreased from 1.02 eV to 0.8 eV with a DCE treatment applied as a function of treatment time, from 6 to 24 hours [38]. T.Y. Kim et al. proposed using varying molar concentrations of DCE solvent for improved control over the doping process in $MoS_2$ [39]. They reported a well-defined correlation between carrier density and the molar concentration of DCE during the 45-minute treatment [39]. While the effects of the DCE treatment on the electrical properties of the aforementioned TMDs have been comprehensively studied and optimised by different research groups, its impact on the optical properties of TMDs remains unclear.

In this study, we systematically investigate the dependence of the optical properties of semiconducting TMDs on both DCE treatment time and layer number. We exfoliated $MoS_2$ flakes with different numbers of layers on PDMS, from monolayer to bulk, and transferred the flakes to fused silica substrates. PL spectroscopy is carried out to investigate the effect on their optical properties with a wide range of DCE treatment durations. To elucidate the energetically favoured Cl-doping mechanisms in $MoS_2$ and the influence of intrinsic sulphur defects on its optical properties, density functional theory (DFT) calculations are performed.

**Methods**

**Experimental:** Bulk single-crystal $MoS_2$ are purchased from HQ Graphene and 2D Semiconductors, which are slightly n-type in their pristine state (see supplementary material) [30]. The Scotch tape and polydimethylsiloxane (PDMS)-assisted mechanical exfoliation method is used to obtain $MoS_2$ flakes with different numbers of layers [40]. Flakes larger than 10 × 10 μm are transferred onto fused silica substrates using a custom-built viscoelastic transfer system. The PL spectra of the flakes are measured using a micro-PL setup equipped with a 500 mm monochromator (Shamrock 500i, Andor), a Si CCD (Newton BEX2-DD, Andor) and a 532 nm excitation laser, before and after DCE treatment. The laser beam is focused to be a spot of ~0.8 μm in diameter with a 100x objective (NA = 0.7). After obtaining PL measurements for pristine flakes, the samples are soaked in the DCE solution for varying durations, ranging from 30 seconds to 24 hours, in a clean room atmosphere. Immediately after DCE treatment, the flakes are dried with $N_2$ gas. Finally, PL measurements are conducted on DCE-treated samples at room temperature. For each sample, the intensity ratio of the PL peaks before and after DCE treatment is calculated to assess the effect of the DCE treatment.

For electrical characterization, the flake is transferred on the $SiO_2$ (300 nm)-on-Si substrate, which is cleaned with acetone (ACE) and isopropyl alcohol (IPA) before the transfer process. The electrodes are patterned on the sample by EBL with a bilayer resist to obtain an undercut structure (MMA/PMMA, Allresist GmbH) and Au/Cr (60 nm/10 nm) is deposited as contact electrodes (Figure S2a). Lift-off process is carried out in warm ACE. The fabricated FET device is mounted onto a ceramic chip holder with wires bonded for electrical characterization. Electrical characterization is carried out on the FET devices before and after 12h of DCE treatment. Output and transfer characterizations are measured using Agilent B2902A source measure unit at room temperature.

**Computational:** DFT calculations are conducted using the Quantum Espresso software package [41,42], to unveil the complex interplay between MoS$_2$ and the Cl-doping process. The projected augmented wave pseudo-potentials [43,44] and the Perdew–Burke–Ernzerhof parametrisation of the generalised gradient approximation exchange-correlation functional are employed [45]. A 6×6×1 K-point mesh within the Gamma-centred Monkhorst-Pack scheme is applied to Brillouin Zone integration with a kinetic energy cut-off of 60 Ry and a density cut-off of 480 Ry. A Gaussian smearing scheme is utilised with a broadening of 0.01 Ry. In the geometry optimisations, the system is allowed to relax fully using a Broyden–Fletcher–Goldfarb–Shanno (BFGS) algorithm along with the total energy threshold of $10^{-5}$ Ry and the force threshold of $10^{-3}$ Ry/Å.

In the calculations, six models are adopted to mimic Cl-doping of the MoS$_2$ by chemical immersion in the experiment: two substitutional and four interstitial doping models. In substitutional doping models of T3 and B3, an isolated Cl atom substitutes for an S atom on either the upper (T3) or the lower S layer (B3) of MoS$_2$. Note that, in the substitutional doping models, metal vacancies are not considered as their formation is energetically highly unfavourable [46]. In the interstitial doping models, an isolated Cl atom is introduced via interstitially insertion at different symmetry-allowed sites: T1, H3, and the van der Waals (vdW) gap within the MoS$_2$ lattice. T1 is a one-fold coordinated site on the topmost Mo atom. H3 is a three-fold coordinated site at the centre of the honeycomb formed by three Mo atoms. VdW gap site is a one-fold coordinated site at the vdW gap of a MoS$_2$ bilayer.

Substitutional doping process (i.e. T3 and B3) mainly consists of two elementary steps: (i) a vacancy formation in MoS$_2$ after the immersion in a solution, leading to the generation of dangling bonds, and (ii) chemical adsorption of Cl on MoS$_2$. The formation energy associated with Cl doping, $E_{form}$ can be computed as a sum of vacancy formation energy, $E_{vac}$, and adsorption energy of Cl, $E_{ads}$.

$$E_{form} = E_{vac} + E_{ads} \qquad (1)$$
$$E_{ads} = E_{doped} + (E_{system} + \eta_{Cl} \times \mu_{Cl}) \qquad (2)$$

where $E_{doped}$ is the total energy of a Cl-doped MoS$_2$ system, $\eta_{Cl}$ is the number of Cl atoms doped in MoS$_2$ and $\mu_{Cl}$ is the chemical potential of a Cl atom in a Cl$_2$ gas-molecule. For the interstitial doping models, the Cl adsorption energy is computed using only Eq. 2.

$$E_{vac} = E_{system} - (E_{pristine} - \eta_s \times \mu_s) \qquad (3)$$

where $E_{system}$ and $E_{pristine}$ are the total energies of a MoS$_2$ sheet with and without vacancy, respectively. $\eta_S$ is the number S atoms removed from the sheet and $\mu_s$ is the chemical potential of a S atom, indicating the energy per atom in α-S bulk.

Following full relaxation, density of states and electronic band calculations are performed on the most energetically favourable model. The models depicted in the figures are visualised using VESTA software [47].

## Results and Discussion

It is well understood that the electronic band structure and bandgap of TMDs depend on the number of layers. Therefore, the number of layers in the TMDs flakes can be precisely determined using the PL peak wavelength/energy [48]. The PL spectrum of each layer is utilised to confirm their number of layers both on the PDMS and fused silica substrate (Figure S1). The peak value for monolayer MoS$_2$ occurs at approximately 655 nm, representing the direct bandgap. The indirect transition peak of bilayer MoS$_2$ starts to appear around 802 nm, while its direct transition peak is around 665 nm. From a monolayer to a thicker flake, the PL intensity dramatically decreases as a function of increased thickness, and the bandgap of MoS$_2$ red shifts up to 1.35 eV, i.e. 921 nm, as shown in the inset of Figure S1. However, distinguishing the indirect bandgaps beyond 7 layers becomes challenging. We refer to more than 7-layer flakes as bulk.

In the literature, T.Y. Kim et al. [39] demonstrated that a 45-minute treatment effectively dopes MoS$_2$, and this finding is used as a reference point for this study. The PL spectra of different layers are characterised before and after a 60-minute DCE treatment (Figure 1). After 60 minutes in the DCE solution, the PL peak intensity of monolayer MoS$_2$ decreases to ~80% of its initial value. Under the same treatment time, the PL peak intensity of the direct transition in bilayer (2L) MoS$_2$ retains 70% ± 15% of its initial intensity, while the indirect transition retains only 18% ± 11% . Repeated 60-minute DCE treatments show a layer-dependent response, where the suppression of the indirect transition diminishes with increasing thickness. The same experiments are repeated with the pristine flakes for 2 minutes of DCE treatment, and it is observed that even with 2 mins treatment time, the PL intensity of the indirect transition for 2L and beyond reduces significantly, while their direct transition intensity remains almost unaffected.

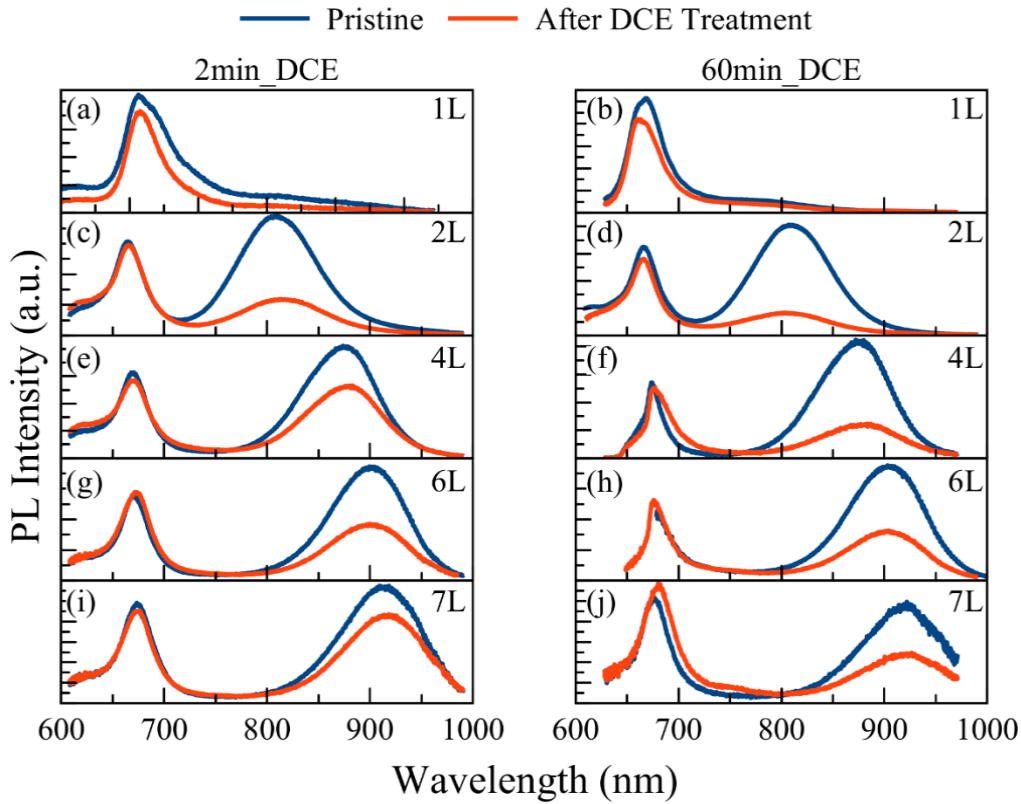

**Figure 1.** The PL spectra of (a, b) 1L, (c, d) 2L, (e, f) 4L, (g, h) 6L and (i, j) 7L $MoS_2$ samples before (dark blue) and after (orange) exposed to DCE solution for 2 minutes (left column) and 60 minutes (right column).

The same DCE treatment is processed on more than a hundred samples with various layer numbers and under different time durations. Figure 2 shows the overview of the PL ratios of DCE-treated to pristine $MoS_2$ samples for both direct and indirect transitions of 1L (direct only), 2L, 4L, 6L, 7L, and bulk, over periods ranging from 30 seconds to 24 hours. The effect of DCE treatment on the PL intensity of 2D $MoS_2$ is assessed for each flake individually by taking the ratio of the peak intensities before and after treatment ($R = I_{dope}/I_{pristine}$). The data points with error bars in Figure 3 and Figure 4 represent the average values of R for all (>100) samples exposed to the same duration in the DCE solution. It is clear that the impact of DCE treatment is not the same on the direct and indirect transitions of $MoS_2$ samples - the indirect transitions are affected more than the direct transitions for all layers. In other words, the indirect transitions in the multilayer $MoS_2$ can be damped, while the direct transition is maintained.

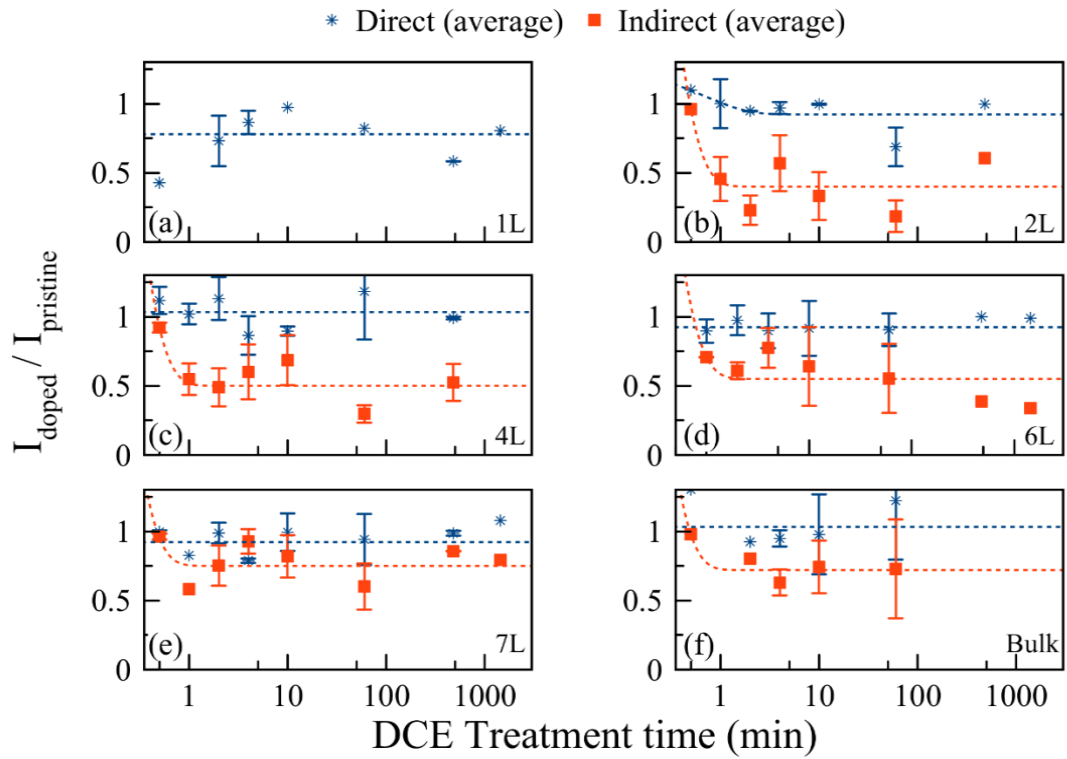

**Figure 2.** Time dependence of the PL intensity ratios before and after DCE treatment for both direct (dark blue) and indirect (orange) transitions of (a) 1L, (b) 2L, (c) 4L, (d) 6L, (e) 7L and (f) Bulk $MoS_2$ under various treatment time in logarithmic scale, ranging from 30 s to 1440 min (24 hours).

The effect of DCE on the optical properties of $MoS_2$, under the same treatment time, also depends on the number of layers. Figure 3 shows the intensity ratio R as a function of the number of layers for different doping durations. The impact of DCE on the PL intensity across all treatment durations reduces as the number of layers increases. The layer dependency of DCE treatment in $MoS_2$ flakes eventually saturated. The dashed line fittings are to highlight the difference in the effect between the direct and indirect transitions. This trend may indicate that the doping process affects only the surface layers.

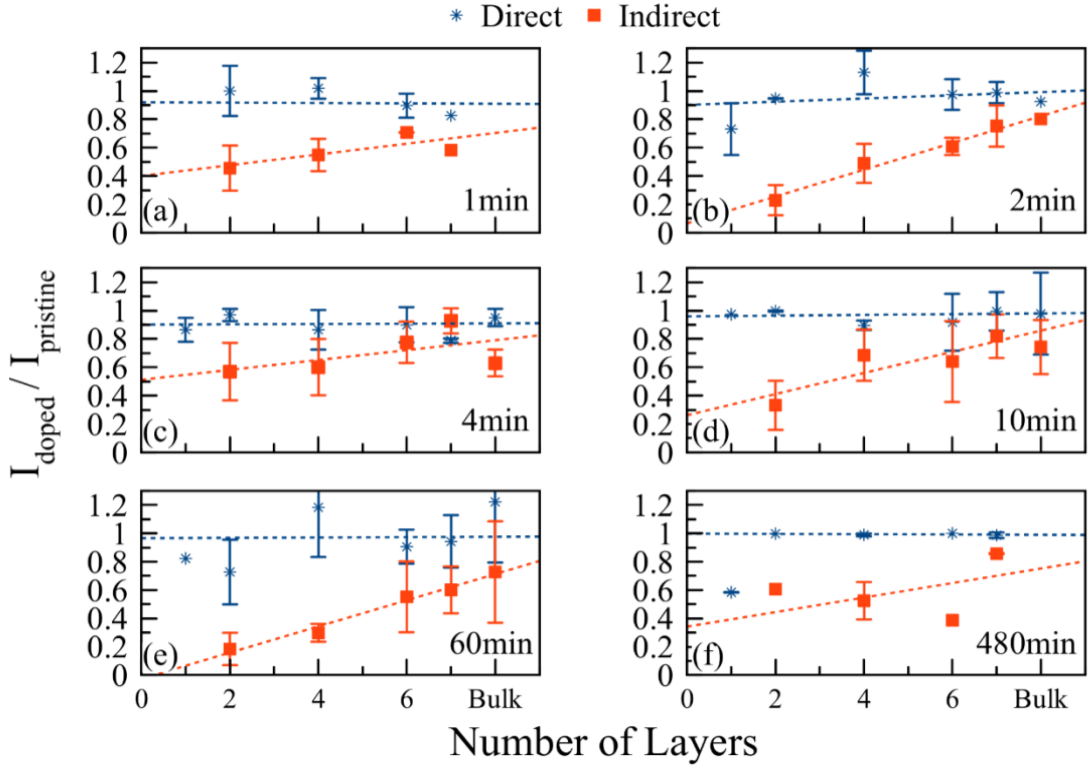

**Figure 3.** 2D MoS$_2$ layer dependence of the PL intensity ratios before and after DCE treatment in both direct (dark blue) and indirect (orange) transitions under (a) 1 min, (b) 2 min, (c) 4 min, (d) 10 min, (e) 60 min and (f) 480 min DCE treatment.

To understand the mechanisms behind the DCE treatment resulting in n-type doping (Figure S2) and its effect on the optical properties of the MoS$_2$, the energetically favourable adsorption mechanisms of Cl and the electronic band structures with density of state (DOS) are calculated in pristine monolayer and bilayer MoS$_2$ flakes for the most possible interaction mechanism scenario. Figure 4 shows the energetic preference of symmetry-permitted doping sites for Cl in MoS$_2$, where the defect formation ($E_{vac}$), Cl binding ($E_{ads}$), and total formation ($E_{form}$) energies, are indicated with orange, blue and green bars. The DFT calculations show that the Cl binding at an interstitial site of a MoS$_2$ lattice is energetically highly endothermic (7 eV for H3, Fig. 4d), while one adsorbed on S-vacancy is exothermic, meaning that Cl binding to defects is an energetically driven process (-1.9 eV for both T3 and B3, Fig. 5b, c). Even with a preceding defect formation in a pristine MoS$_2$, the total energy required for substitutional doping of Cl for an S atom is still lower (1.8 eV for both T3 and B3) than those for interstitial doping (H3 and vdW gap, Fig. 4e). When comparing the total formation energies, Cl adsorption on a pristine surface is energetically most favourable (0.9 eV for T1) – which is the essential step to attract Cl to the surface (Fig. 4a). This is followed by the energies for the replacement of a surface S atom by a Cl atom in T3 and B3 models. Cl doping at the vdW gap **(VDG in Fig. 4e)**, is energetically more expensive than substitutional doping, while cheaper than the interstitial doping.

These results indicate that the most likely scenario is that the Cl atoms are filling the S-vacancies or replacing S atoms in MoS$_2$ during the doping process. Moreover, the DFT results suggest that, depending on the local chemical environment and doping conditions, it is also likely for Cl atoms to migrate at the vdW gap in the MoS$_2$ multilayers – this indeed increases the interlayer spacing while reducing the interlayer coupling, which can alter the electronic band structure of MoS$_2$.

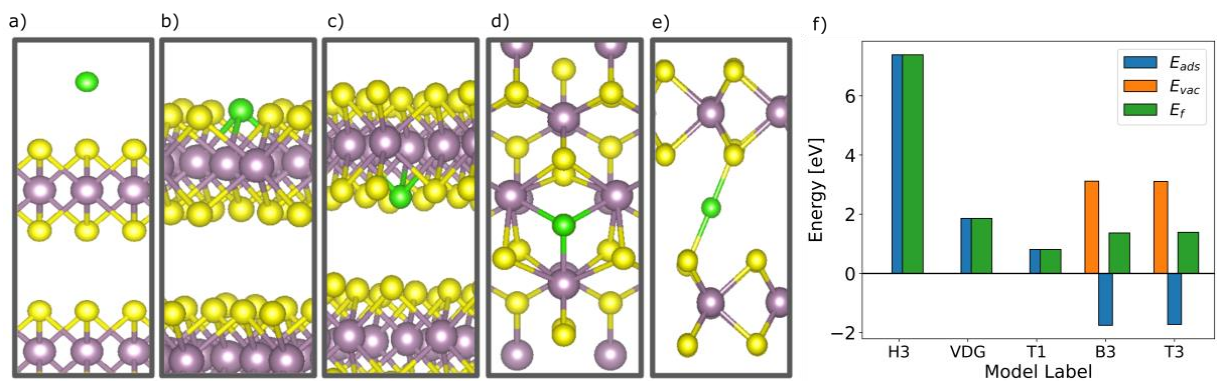

**Figure 4.** Optimised configurations of a MoS$_2$ bilayer with Cl (a) adsorbed on the surface (T1), (b) replacing a S atom on the upper S (T3) and (c) the lower S layer of MoS$_2$ (B3), (d) inserted at the centre of honeycomb neighbouring by three Mo atoms (H3) and (e) inserted at the vdW gap. (f) A graph depicting S-vacancy formation energy $E_f$, which is the sum of $E_{vac}$ and $E_{ads}$, along with $E_{vac}$ and $E_{ads}$.

The band structure analysis in Figure 5 highlights the effect of treatment on the electronic band structure of materials. In the presence of substitutional doping of Cl in a bilayer MoS$_2$, the bandgap still exhibits indirect bandgap characteristics with a narrower bandgap (Fig 5c, d) compared to the pristine layer (Fig 5a, b) while the insertion of Cl at the vdW gap, i.e. VDG, causes less dispersive (flatter) bands resulting in both lowering the bandgap and transition from indirect to direct bandgap and (Fig 5e).

Density of State (DOS) calculations give additional support to our hypothesis and reveal the effect of dopant contribution near the Fermi level energy states (Figure 5f-g-h). This leads to an increase in electrical conductivity. The emergence of an additional nearly flat band around the Fermi energy level indicates strong electronic localisation due to the VDG dopant (Figure 5e, h). Also, the indirect bandgap feature significantly diminishes upon vdW gap doping, compared to the substitutional doping in T3 and B3 models (Figure 5c, d, g). This suggests, while both substitutional and interstitial doping are likely to suppress the indirect bandgap feature as observed in experiments, the vdW gap type doping is likely to induce the bandgap transition. All these results together indicate that Cl-doping is a candidate mechanism for engineering the bandgap.

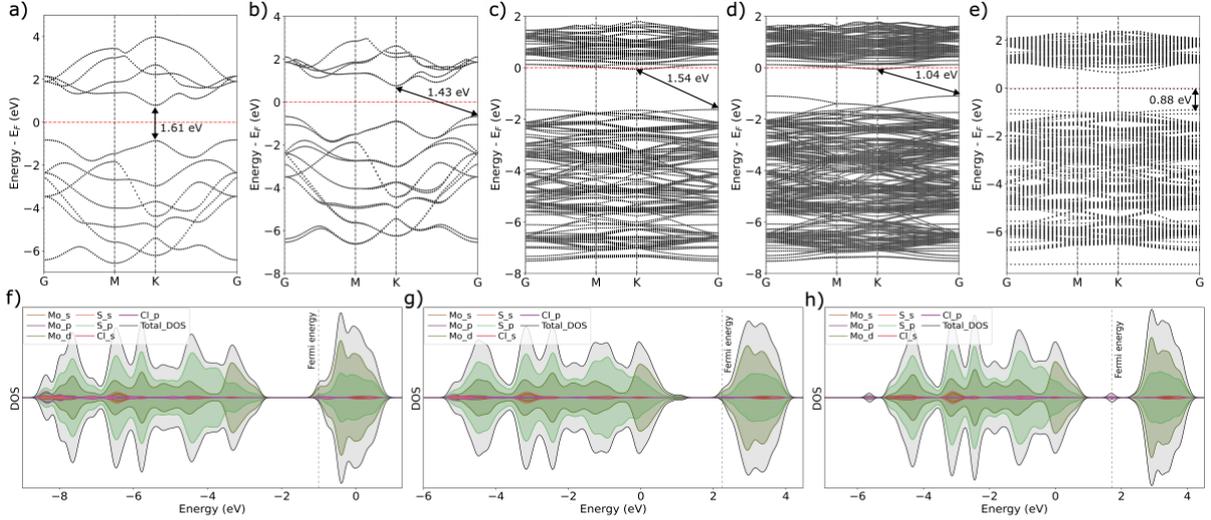

**Figure 5.** DFT computed electronic bandgap structures of a pristine (a) monolayer and (b) bilayer $MoS_2$; and a bilayer $MoS_2$ with Cl doped on (c) T3, (d) B3 and (e) at the vdW gap. Density of state (DOS) calculations of (f) a pristine bilayer, Cl-doped on (g) T3 and (h) at the vdW gap.

Chlorine (Cl) atoms are found to settle at defect sites near the $MoS_2$ indirect band level. This facilitates non-radiative relaxation from the conduction to the valence band, thereby quenching the PL intensity of the indirect transition. We attribute the unaffected direct transition to its occurrence solely at the K-point, which is relatively isolated in momentum space and thus less sensitive to defect-induced scattering or trapping. Defects arising from Cl-doping introduce localised states that primarily affect extended states across the band structure, leaving the highly localised direct transition largely unaffected.

**Conclusion**

In this study, we systematically investigated the impact of chlorine-based chemical treatment using 1,2-dichloroethane (DCE) on the optical properties of two-dimensional $MoS_2$ with varying layer numbers and treatment durations. Through comprehensive PL measurements, a suppression in the indirect transition peak intensity is observed, even after short DCE exposures (i.e., 2 minutes), while the direct transition remains largely unaffected across all layer numbers.

This effect becomes more prominent with increased treatment duration, and diminishes with increased layer thickness, which indicates a layer- and time-dependent doping behaviour. Density functional theory calculations revealed that chlorine atoms preferentially occupy sulphur vacancy sites in $MoS_2$, forming energetically favourable configurations. These Cl dopants introduce mid-gap states near the indirect band edge, promoting non-radiative recombination pathways that selectively reduce the indirect PL transitions. This insight suggests a transition from an indirect to direct bandgap character in multilayer $MoS_2$, effectively enabling bandgap engineering through surface chemical doping. It is demonstrated that DCE treatment is a tuneable post-growth strategy not only for n-type doping but also for manipulating the optical response of $MoS_2$. The ability to selectively suppress indirect transitions while preserving direct transitions opens new possibilities for enhancing the performance of TMD-based optoelectronic devices, especially where direct-gap emission is crucial.

**Acknowledgements**

This study was supported by the Scientific Research Projects Executive Secretary of Istanbul University, project numbers FBA-2023-39412 and FYL-2023-39742, and the Scientific and Technological Research Council of Turkey (TÜBİTAK) 1001 – Scientific and Technological Research Projects Support Program, project number 121F169. Y.W. acknowledges a Research Fellowship awarded by the Royal Academy of Engineering RF/201718/17131 and an EPSRC grant EP/V047663/1. The authors would like to thank Dr. Alexander Armstrong and Prof. Keith McKenna for their valuable discussions on DFT.

**Author Contributions**

Y.K.B., E.H. and F.S. fabricated the samples, performed the steady-state photoluminescence measurements; C.S. performed the DFT calculations and supervised by N.N; Y.K.B., N.N., A.E., Y.W. and F.S., analysed the results; A.E., Y.W and F.S. managed various aspects and funded the project; Y.K.B., and F.S. wrote the manuscript with contributions from all co-authors; Y.W. and F.S. oversaw the entire project. All authors have read and agreed to the published version of the manuscript.

**Data and code availability**

The authors declare that all the data and code supporting the findings of this study are available within the article, or upon request from the corresponding author.

**Corresponding Author**

* Yue Wang: School of Physics, Engineering and Technology, University of York, York, YO10 5DD, United Kingdom, orcid.org/0000-0002-2482-005X, email: yue.wang@york.ac.uk

* Fahrettin Sarcan: Department of Physics, Faculty of Science, Istanbul University, Vezneciler, 34134, Istanbul, Turkey, orcid.org/0000-0002-8860-4321, email: fahrettin.sarcan@istanbul.edu.tr

# Supplemental Material for

# Chemical Treatment-Induced Indirect-to-Direct Bandgap Transition in MoS$_2$: Impact on Optical Properties

Yusuf Kerem Bostan[1], Elanur Hut[1], Cem Sanga[2], Nadire Nayir[2,3], Ayse Erol[1], Yue Wang[4] *, and Fahrettin Sarcan[1,4] *

[1]Department of Physics, Faculty of Science, Istanbul University, Vezneciler, Istanbul, 34134, Turkey

[2]Department of Physics Engineering, Istanbul Technical University, Maslak, Istanbul, 34467, Turkey

[3]Paul-Drude-Institute for Solid State Electronics, Leibniz Institute within Forschungsverbund Berlin eV., Hausvogteiplatz 5-7, 10117 Berlin, Germany

[4]School of Physics, Engineering and Technology, University of York, York, YO10 5DD, United Kingdom

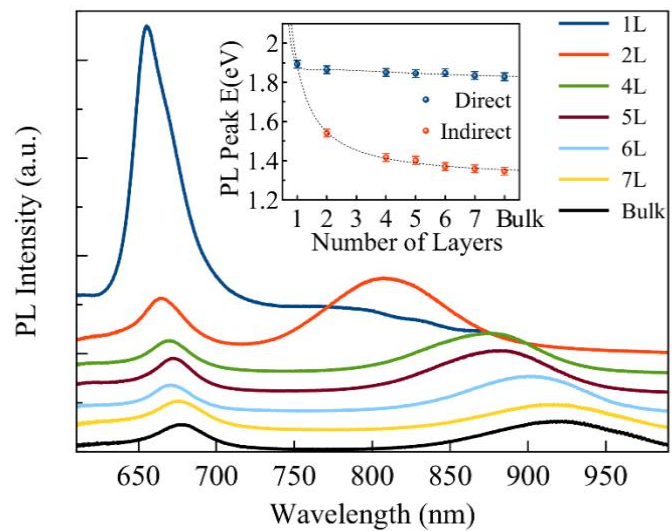

**Figure S1**. PL spectra of MoS$_2$ flakes with different numbers of layers, from monolayer to bulk. The inset shows the bandgap energy of each layer extracted from PL peaks.

The output characteristics shows the drain current increased up to −1.5 µA under a 1 V negative drain bias across 3 µm channel length at 0 V gate voltage (Figure S2b). According to the transfer characteristics of the 3 µm channel device, $I_{on}/I_{off}$ ratio of the pristine FET is 2 orders of magnitude and is improved to 6 orders of magnitude after 12 hours of treatment (Figure S2c). The pristine sample exhibits n-type characteristics, which is enhanced after 12 hours in DCE solution.

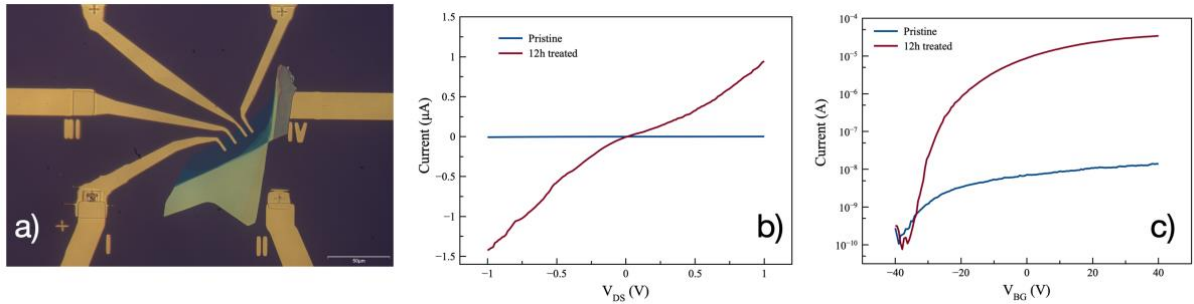

**Figure S2. a)** Optical microscope images of few layer MoS$_2$ based FET device, **b)** output and **c)** transfers characteristic of the devices before and after DCE treatment.